Mid-infrared assisted water electrolysis as a method to substantially push down the overpotential of the oxygen evolution reaction

Klara Rüwe and Helmut Schäfer[*]

University of Osnabrück, The Electrochemical Energy and Catalysis Group,

Barbarastrasse 7, 49076 Osnabrück, Germany

E-mail: helmut.schaefer@uni-osnabrueck.de

**Abstract**

Photocatalytic- and Photoelectrochemical water splitting is currently performed using radiation sources with wavelengths < 1000 nm, *i.e.* in the near infrared range (NIR). The fact that water has a broad absorption band, which lies at wavenumbers between 3000 and 3700 $cm^{-1}$ (stretching mode) and 1650 $cm^{-1}$ (bending mode), has not been taken into account so far. We irradiated the steel anode with a mid-infrared LED ($\lambda$=3300 nm) while water electrolysis was performed in pH 7-corrected phosphate buffer solution. A significant shift (270 mV) of the cyclic voltammetry (CV) curve towards lower potential values was obtained when the radiation source was on. To the best of our knowledge, this is the first report describing the use of a mid-infrared radiation source to increase the efficiency of water electrolysis.

**Introduction**

Electrocatalytic-, photocatalytic- and photoelectrochemical splitting of water into its cleavage products when sourced by renewable energy represents a promising, $CO_2$ footprint free solar to fuel conversion route [1, 2, 3, 4].

The efficiency of water electrolysis (electrocatalytically initiated water) stands and falls with the cell voltage which is the sum of the thermodynamic decomposition voltage (1.23 V) plus the overvoltage occurring at both electrodes [5]. Therefore, high overpotentials, which can be attributed to both half-cell reactions, the hydrogen evolution reaction, as well as the oxygen evolution reaction, still represent a hurdle to be overcome. Currently, the optimization of electrocatalytically initiated water splitting is based on the improvement of the electrode materials [6, 7, 8], the electrolyte [9] respectively.

To the best of our knowledge, most photocatalytic or photoelectrochemical approaches to assist in the splitting of water molecules into their fission products utilize visible (400–800 nm) or near-infrared (800–1000 nm) radiation sources), is therefore operated with diodes that work at λ =400-1000 nm[10, 11].

It is known that due to the excitation of the stretching vibration and the bending vibration, water has broad absorption bands that are in the infrared spectral range[12, 13], for example at wavelengths around 2800 nm and 6000 nm. Irradiation of water with diode lasers of appropriate wavelength should weaken the intramolecular O-H bond and lead to a significant reduction in overpotentials. We conducted experiments with a mid-infrared LED with a wavelength of 3300 nm and indeed achieved a significant reduction in the overall cell voltage. In particular, it has been found that the potential of the anode at which oxygen evolution begins is significantly reduced when irradiated with mid-IR radiation. To the best of our knowledge, most of the photocatalytic approaches described so far to improve water electrolysis (photoelectrochemical water splitting) only use radiation sources with significantly shorter wavelengths.

**Results and discussion**

In order to show the effect of using Mid-IR radiation source on the water splitting capabilities of a steel anode (steel X20CoCrWMo10-9, see experimental section), water electrolysis was realized by the simplest approach imaginable from an experimental point of view upon using a glass beaker (single compartment conditions) and a three- electrode configuration consisting of a steel anode, a Pt counter electrode and a reversible hydrogen electrode. Water splitting was carried out (for comparison) in a pH 7 corrected mixture of 0.1 molar $KH_2PO_4$ and 0.1 molar $K_2HPO_4$ solution first without using radiation. The CV curve can be taken from Figure 1 (black curve). Then the LED (4 x 4 mm in size) was positioned directly in front of the steel anode electrode (between reference and the anode) and the experiment was repeated. The CV curve can be taken from Figure 1 (blue curve). The substantial shift of the CV curve towards lower potentials can be clearly seen. The potential at which oxygen evolution begins (onset of OER) defined as the potential (in V vs. RHE) at which the current density exceeds 100 µA/cm$^2$ was reduced by 270 mV, i.e. the CV curve shifted from 1.57 V vs. RHE (without irradiation) to 1.3 V vs. RHE (with irradiation). It should be mentioned at this point that non-activated X20CoCrWMo10-9 steel has been used as electrode material which is known to have a rather slow OER activity (See sample Co in our previous publication[6]). Therefore, an onset potential of 1.3 V vs. RHE determined in pH 7 medium has to be seen as a fantastic value achieved with a rather inactive anode material. In addition, it turned out that the overall cell voltage was also reduced. These findings impressively prove the suitability of this Mid-IR-based approach to improve the water splitting capabilities of known steel-based electrode materials. To eliminate eventually occurring charging of thin oxide layers whilst recording CV's (that were formed on the steel surface upon positive potentials at the anode) we repeated measurements at lower scan rate (10 mV/s). However, it turned out that this did not influence the gap between CV curves derived from measurements achieved upon irradiation, without using Mid-IR source respectively.

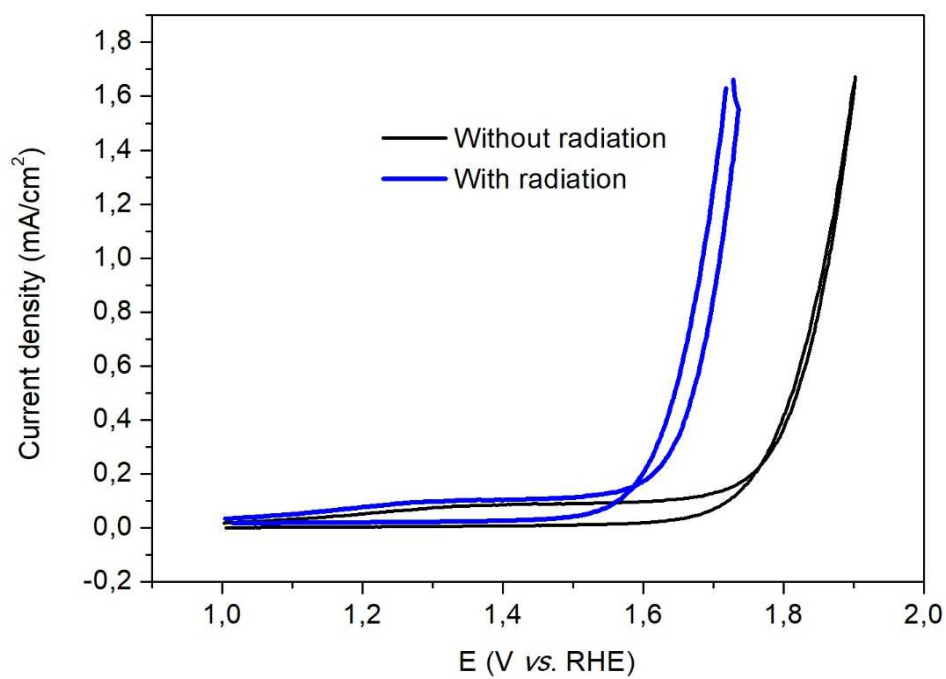

Figure 1. Results from cyclic voltametric measurements. Electrolyte: pH 7 corrected phosphate buffer solution. Anode: untreated X20CoCrWMo10-9 steel, electrode area: 2 cm$^2$. Scan rate: 20 mV/s. The CV curve (with radiation) was recorded whilst the LED was switched on.

**Conclusions**

We have shown upon rather simple electrochemical measurements (quasi-steady state conditions) that the exploitation of Mid-IR irradiation of the anode is a reasonable and effective method to substantially improve the water electrolysis performance.

**Experimental part:**

**Electrochemical measurements** A three-electrode set-up was used for all electrochemical measurements. The working electrode (WE) consisting of electro-activated AISI 304 steel (X20CoCrWMo10-9) (sample Co) was prepared as described in our previous publication[6]. A Pt wire electrode (4x5 cm geometric area) was exploited as counter electrode and a reversible hydrogen electrode served as reference electrode (RHE, HydroFlex, Gaskatel Gesellschaft für Gassysteme durch Katalyse und Elektrochemie mbH. D-34127 Kassel, Germany) was utilized as the reference standard, therefore all voltages are quoted against this reference electrode (RE).

For all measurements the RE was placed between the working and the CE. The measurements were performed in a pH 7 corrected 0.1 M $KH_2PO_4/K_2HPO_4$ solution. The pH 7 corrected 0.1 M KH2PO4/K2HPO4 solution was prepared as follows: Aqueous solutions of 0.1 M $K_2HPO_4$ and $KH_2PO_4$ (VWR, Darmstadt, Germany) were mixed until the resulting solution reached a pH value of 7.0. The distance between the WE and the RE was adjusted to 1 mm and the distance between the RE and the CE was adjusted to 4-5 mm. Irradiation of the anode was realized upon usage of a NanoPlus MIR LED; NanoPlus D-97218 Gerbrunn, Germany). The distance between anode and LED was adjusted to < 1mm. All electrochemical data were recorded digitally using a Potentiostat Keithley Tektronix 2460 SourceMeter®.

Cyclic voltammograms (CV) were recorded in 100 mL of electrolyte in a 150 mL glass beaker under stirring (180 r $min^{-1}$) using a magnetic stirrer (15 mm stirring bar). The scan rate was set to 20 mV $s^{-1}$ and the step size was 2 mV. The potential was cyclically varied between 1- and 1,9 V *vs.* RHE. No IR compensation was performed whilst recording the CV plots.